\documentstyle[aps]{revtex}
\begin{document}
\title {Proton Particle-Neutron Hole States in $^{132}$Sb \\
with a Realistic Interaction}
\author{F. Andreozzi, L. Coraggio, A. Covello, A. Gargano, T. T. S. Kuo 
and A. Porrino}
\address{Dipartimento di Scienze Fisiche, Universit\`a 
di Napoli Federico II, \\ and Istituto Nazionale di Fisica Nucleare, \\
Complesso Universitario di Monte  S. Angelo, Via Cintia - I-80126 Napoli, 
Italy}
\date{\today}
\maketitle
\begin{abstract} 
The structure of the particle-hole nucleus $^{132}$Sb provides a direct
source of information on the effective neutron-proton interaction
in the $^{132}$Sn region. We have performed a shell-model calculation
for this nucleus using a realistic effective interaction derived from the 
Bonn A nucleon-nucleon potential. The results are in very good agreement
with the experimental data evidencing the reliability of our 
effective interaction matrix elements both with isospin $T=1$ and
$T=0$.

\end{abstract}

\draft
\pacs{21.60.Cs,21.30.Fe,27.60.+j}

A fundamental problem in nuclear structure theory is to understand the
properties of complex nuclei in terms of the nucleon-nucleon ($NN$)
interaction. Since the initial work of Kuo and Brown \cite{KB}, who
derived an {\it s-d} shell effective interaction from the Hamada-Johnston
potential \cite{HJ}, there has been much progress in this field.
On the one hand, the many-body methods for calculating the matrix
elements of the effective interaction have been largely improved
(a concise review of the state of the art is given in
Ref. \cite{Kuo96}). On the other hand, more realistic {\it NN} potentials
have been developed which are able to reproduce quite accurately all
the known $NN$ data. A comprehensive account of these developments
through 1993 is given in the review paper by Machleidt and Li
\cite{Machl94}.

Motivated by these improvements, in recent years there has been a
revival of interest in nuclear structure calculations with realistic
effective interactions. As a result, calculations of this kind, which
in earlier studies had been mainly confined to light nuclei,
have been recently extended to medium- and heavy-mass nuclei. Until now,
however, attention has been focused on identical particle systems, 
in particular on Sn isotopes and $N=82$ isotones 
\cite{Eng95,Andr96,Andr97,Holt97}.
In our own studies \cite{Andr96,Andr97} (hereafter referred to as I and II)
we considered the $^{100}$Sn neighbors going from $^{102}$Sn to
$^{105}$Sn while for the $N=82$ isotones we were concerned with the
$^{132}$Sn neighbors with two and three valence protons. In both works
we performed shell-model calculations using a realistic effective
interaction derived from the meson-theoretic Bonn A potential
\cite{Machl87}. 

The motivation for the theoretical study of nuclei around doubly magic
$^{100}$Sn and $^{132}$Sn lies in the fact that they provide the best testing 
ground for the basic ingredients of shell-model calculations in the
100-150 mass region, especially as regards the matrix elements 
of the effective $NN$ interaction.
These nuclei, however, lie well away from the valley of stability and
until recently experimental information on their spectroscopic properties
was very scanty. The advent of large multidetector $\gamma$-ray arrays
and radioactive ion beam facilities is now making more and more
accessible their study. In this context, it is worthwhile to mention 
that in I we could only predict the spectra
of $^{102}$Sn and $^{103}$Sn since  no experimental data 
for these two very neutron deficient nuclei were available at that time. 
After submission of this paper the identification of three excited 
states of $^{102}$Sn in an in-beam $\gamma$-ray spectroscopic experiment 
was reported \cite{Lipo96}. As is shown in Ref. \cite{Cov97}, 
our predictions turned out to be in remarkably good agreement with 
the experimental results. Similarly, our study II of $^{134}$Te and 
$^{135}$I has largely benefited from the results of very recent 
experimental work \cite{Omtvedt95,Zhang96}.

The very good agreement between theory and experiment achieved in
I and II evidences the reliability of our $T=1$ effective interaction
in the 100-150 mass region and makes apparent the motivation for the 
present study of the doubly odd nucleus $^{132}$Sb. 
This nucleus, which has a single proton
outside the $Z=50$ closed shell and a single neutron hole in the closed
$N=82$ shell, offers the opportunity to test the $T=0$ matrix elements
of the effective interaction. Experimental information on
$^{132}$Sb was earlier provided by the works of Refs. \cite{Kerek72}
and \cite{Stone89}. Recently, the detailed study of Ref. \cite{Mach95}
has significantly improved our knowledge of its spectroscopic
properties.

In our shell-model study of $^{132}$Sb we consider the doubly magic $^{100}$Sn
as a closed core and treat the odd proton and the remaining 31 neutrons
as valence particles. The model space consists of the five single-particle
(s.p.) orbits $0g_{7/2}$, $1d_{5/2}$, $1d_{3/2}$, $2s_{1/2}$, and $0h_{11/2}$,
and use is made of a two-body effective interaction  derived
from the Bonn A nucleon-nucleon potential. This was obtained using a $G$-matrix
formalism, including renormalizations from both core polarization and
folded diagrams. We have chosen
the Pauli exclusion operator $Q_2$ in the $G$-matrix equation,
$$G(\omega)=V+V Q_2 {{1} \over
{\omega-Q_2TQ_2}} Q_2G(\omega), \eqno (1)$$
as specified by ($n_1, n_2, n_3$) = (11, 21, 45) \cite{Kuo96}. Here $V$
represents the $NN$ potential, $T$ denotes the two-nucleon kinetic energy,
and $\omega$ is the so-called starting energy. We employ a matrix inversion
method to calculate the above $G$ matrix in an essentially 
exact way.
The effective interaction, which is energy
independent, can be schematically written in operator form as
$$V_{eff} = \hat{Q} - \hat{Q}' \int \hat{Q} + \hat{Q}' \int \hat{Q} \int
\hat{Q} - \hat{Q}' \int \hat{Q} \int \hat{Q} \int \hat{Q}\, \cdots \; ,
\eqno (2) $$
where $\hat{Q}$ and $\hat{Q'}$ represent the $\hat{Q}$-box, composed of
irreducible valence-linked diagrams, and the integral sign represents a
generalized folding operation. We take the $\hat{Q}$-box to be
composed of $G$-matrix diagrams through second-order in $G$.
The shell-model oscillator parameter used by us is $\hbar \omega$ = 8 MeV.
It should be noted that the $T=1$ matrix elements of the effective
interaction used in the present calculation are just the same as those
used in I.
A detailed description of our derivation of $V_{\rm eff}$
can be found in Ref. \cite{Kuo96}. 

As regards the s.p. energies, we assume them to be the same for neutrons 
and protons. Our adopted values (in MeV) are: $\epsilon_{d_{5/2}}=0$,
$\epsilon_{g_{7/2}}=0.20$, $\epsilon_{s_{1/2}}=1.72$, $\epsilon_{d_{3/2}}=1.88$,
and $\epsilon_{h_{11/2}}=2.70$. As compared to the set of s.p. energies used 
in I, only $\epsilon_{s_{1/2}}$ and $\epsilon_{d_{3/2}}$ have been
modified. It should be noted that their position played a minor role 
in our earlier study I of the light Sn isotopes.
Here we have kept constant the spacing between these two levels while
shifting down their position by about 0.5 MeV. We have found that this change
is essential to satisfactorily reproduce the experimental $1_2^+$ state in 
$^{132}$Sb and to place the calculated negative-parity states in the right 
energy range.
In this context, it is worth mentioning that we have also calculated
the spectra of $^{133}$Sb and $^{131}$Sn making use of the above set
of s.p. energies. These nuclei have a single proton outside the
$Z=50$ closed shell and a single neutron hole in the $N=82$ closed
shell, respectively. It turns out that the corresponding
experimental spectra are reproduced within about 200 keV, a larger 
discrepancy (780 keV) occurring only for the ${7 \over 2}^+$ state in 
$^{131}$Sn. In this regard, it is interesting to note that our calculated
position of the ${3 \over 2}^+$ single proton level in $^{133}$Sb (2.4 MeV)
is in very good agreement with the new value (2.439 MeV) very recently
provided by the high-sensitivity $\gamma$ spectroscopic measurement
of Ref. \cite{Sanchez}.

Let us now come to the results of our calculation. The experimental and
theoretical spectra of $^{132}$Sb are compared in Fig. 1, where all the
observed levels are reported. In the calculated spectrum all levels 
up to 1.4 MeV excitation energy are included while in the higher energy
region only the $1^+_2$ and $3^-_2$ states are reported. It should be 
noted that the
nature of the presently available experimental information is quite
different for positive- and negative-parity levels. In fact, while the
spin-parity assignments to the former have been clarified by the study of Ref.
\cite{Mach95}, this is not the case for the latter. More precisely,
the excitation energy of the $8^-$ state is not known (the work of
Ref. \cite{Stone89} places it between 150 and 250 keV) and the three other
observed negative-parity states have not received firm spin assignments.
We find that the first excited $8^-$ state lies at 126 keV while the
$6^-_1$, $5^-_1$, $7^-_1$, $3^-_1$, and $4^-_1$ states are grouped 
in a very small energy interval (from 214 to 380 keV). As a consequence,
any attempt to establish a one-to-one correspondence between the
observed levels and those predicted by our calculation could be
misleading. The relevant outcome of our calculation is that the
above states, which all arise from the $\pi g_{7/2} \nu h_{11/2}$
configuration, lie very close in energy and are well separated from the
other two members of the multiplet, i.e. the $9^-$ and $2^-$
states, which are predicted at 1.01 and 1.42 MeV,
respectively. A similar behavior is also predicted for the
$\pi d_{5/2} \nu h^{-1}_{11 /2}$ multiplet. In fact, the calculated 
energies of the $7^-$, $6^-$, $5^-$, and $4^-$ states belonging to
this configuration go from 0.82 to 0.97 MeV while the highest- and
lowest-spin members ($8^-$ and $3^-$) lie at 1.12 and 1.56 MeV,
respectively.

From Fig. 1 we see that the experimental excitation energies of the
positive-parity states are remarkably well reproduced by the theory,
the largest discrepancy being 77 keV for the $5^+_1$ state. The value of the
rms deviation $\sigma$ \cite{sigma} is only 32 keV. In earlier
works (\cite{Kerek72,Stone89} and references therein) the first four 
observed positive-parity states were attributed
to the $\pi g_{7/2} \nu d^{-1}_{3/2}$ configuration and the $3^{(+)}$
level at 529 keV to the $\pi g_{7/2} \nu s^{-1}_{1/2}$ configuration.
The $2^+_2$ and the $1^+_1$ states were interpreted as members of
the $\pi d_{5/2} \nu d^{-1}_{3/2}$ multiplet, and the $1^+_2$ state
as a member of the $\pi g_{7/2} \nu d^{-1}_{5/2}$ multiplet.
Our calculated wave functions, however, are not
really pure. We find, in fact, that the contribution coming from configurations
other than the dominant one is particularly significant for most of the
calculated states. More precisely, the percentage of these components
ranges from 2\% to 20\%. Below 1.4 MeV we also predict the existence of the 
non observed members of the $\pi g_{7/2} \nu s^{-1}_{1/2}$
and $\pi d_{5/2} \nu d^{-1}_{3/2}$ multiplets. They are the $4^+_2$
and the $3^+_3$ and $4^+_3$ states, respectively.

In Table I we compare the experimental electromagnetic transition probabilities
in $^{132}$Sb with the calculated ones. The $E2$ transitions have been
calculated using an effective proton charge $e_p^{\rm eff}=1.55e$, which is
the same as that adopted in II for the $N=82$ isotones. No effective charge
has been attributed to the neutron hole. As regards the magnetic transitions,
for the proton gyromagnetic factors we have adopted the values
$g_s^{\rm eff}(p)=4.465$ and $g_l^{\rm eff}(p)=1.55$ while $g_s^{\rm free}(n)$
and $ g_l^{\rm free}(n)$ have been used for the neutron hole.
As we see from Table I, most of the experimental data are
affected by large errors and for several transition rates only an upper
or lower limit is available from experiment. In view of this, the
agreement between theory and experiment for the $B(E2)$'s can be
considered quite satisfactory. In fact, our calculated values lie all
but two within the limits set by experiment. Concerning the $B(M1)$'s,
the five transitions for which a definite value is available from
experiment are not reproduced by our calculation within the error
bars. In this connection, it is to be noted that, should one ignore
configuration mixing, the three $M1$ transitions $2^+_2 \rightarrow 3^+_2$,
$2^+_2 \rightarrow 2^+_1$ and $2^+_2 \rightarrow 3^+_1$ would be forbidden
while the same value would be predicted for the $3^+_1 \rightarrow 4^+_1$
and $2^+_1 \rightarrow 3^+_1$ transitions. It therefore appears that 
our calculation produces a configuration mixing which goes in the right 
direction. It is clear, however, that for a stringent test of the
wave functions additional and more accurate experimental information
is needed. 

At this point, it should be mentioned that several calculations
have been previously performed to predict the structure of $^{132}$Sb.
A comprehensive account of the earlier works including references
can be found in \cite{Stone89}. We only comment here on the most recent 
calculation of Ref. \cite{Erok94}, which has been carried out by using the
random-phase approximation with a phenomenological interaction.
The results of this calculation are discussed in detail in Ref.
\cite{Mach95}. 

As regards the positive-parity spectrum, the overall agreement with
experiment obtained in Ref. \cite{Erok94} for
the first five excited states is about the same as ours, while for the two 
higher-lying $1^+$ states our results are significantly better. 
As for the electromagnetic transition rates, most of the theoretical
values reported in Ref. \cite{Mach95} are also close to ours. In the
few cases where there are significant discrepancies no definite
conclusion can be drawn owing to the large experimental uncertainties.
One notable exception is the $B(E2; 3^+_1 \rightarrow 4^+_2)$ for which
a value of $36 \pm 11$ e$^2$fm$^4$ has been measured \cite{Mach95}.
Our calculated value, 42.1 e$^2$ fm$^4$, falls within the error bar,
while the calculation of Ref. \cite{Mach95} overstimates the experimental
value by a factor of two. This overestimate, however, can be
mainly attributed to the large value of the neutron effective
charge ($0.9 e$) adopted in Ref. \cite{Erok94}.

From the above comparison we see that our calculation seems to lead to
an even better agreement with experiment than that produced by
the sophisticated approach of Ref. \cite{Erok94}. We would like
to emphasize that this has been achieved by using a realistic
effective interaction with a sound meson-theoretic basis.

In summary, we have shown that our effective interaction derived from
the Bonn A nucleon-nucleon potential leads to a very good description
of the proton particle-neutron hole nucleus $^{132}$Sb. To our 
knowledge, this is the first test of the $T=0$ matrix elements
of this interaction in the 100-150 mass region. A particular feature
of the Bonn A potential relevant to the present study is its
weak tensor force. In fact, in earlier works using different $NN$ potentials
it turned out that not enough attraction was provided by the
calculated matrix elements of the $T=0$ effective interaction,
which has a stronger dependence on the tensor force strength
than the $T=1$ interaction (a detailed discussion of this
important point including references is given in Ref.
\cite{Jiang92}). Our results indicate that the Bonn A potential
is quite suitable for use in shell-model
studies of nuclei with both like and unlike nucleons in valence shells.

\acknowledgments
This work was supported in part by the Italian Ministero dell'Universit\`a
e della Ricerca Scientifica e Tecnologica (MURST) and by the U.S. DOE Grant
No. DE-FG02-88ER40388. We thank H. Mach for valuable comments.

\begin{figure}
\caption{Experimental and calculated spectrum of $^{132}$Sb.}
\label{fig.1}
\end{figure}

\begin{table}
\caption {Calculated and experimental transition rates in 
$^{132}$Sb. The experimental data are from [16].}
\begin{tabular}{lldld}
$J^\pi_i \rightarrow J^\pi_f$ & $B(M1)_{\rm{expt}}$ & 
$B(M1)_{\rm {calc}}$ 
& $B(E2)_{\rm {expt}}$ & $B(E2)_{\rm {calc}}$ \\
&($10^{-3}\mu_{N}^{2}$) &($10^{-3}\mu_{N}^{2}$) & (e$^{2}$ fm$^{4}$) & 
(e$^{2}$ fm$^{4}$) \\
\tableline
$3^+_1 \rightarrow 4^+_1$ & 2.02 $\pm$ 0.03 &12 & 36 $\pm$ 11 & 42  \\
$2^+_1 \rightarrow 3^+_1$ & 61 $\pm$ 7 & 35 & 76 $\pm$ 76 & 53  \\
$\phantom{2^+_1 \rightarrow} ~4^+_1$ &  & & $ < $ 26 & 5.8  \\
$3^+_2 \rightarrow 2^+_1$ & & 5.0 &  & 6.1 \\
$\phantom{3^+_2 \rightarrow} ~3^+_1$ &  $ > $ 3.4   & 257 &  & 3.4 \\
$\phantom{3^+_2 \rightarrow} ~4^+_1$ & $ > $ 17.5   & 409 &  & 18  \\
$2^+_2 \rightarrow 3^+_2$ & 4.8 $\pm$ 2.6 & 1.1 & ${1.1^{+15.4}_{-1.1}}$
& 0.85 \\
$\phantom{2^+_2 \rightarrow} ~2^+_1$ & 2.2 $\pm$ 1.7  & 0.0001 & 37 $\pm$ 30  
& 4.4  \\
$\phantom{2^+_2 \rightarrow} ~3^+_1$ & 10 $\pm$ 5  & 0.90 & 36 $\pm$ 23   
& 5.4  \\
$\phantom{2^+_2 \rightarrow} ~4^+_1$ & &  & 8.4 $\pm$ 4.5  & 9.1  \\
$1^+_1 \rightarrow 2^+_2$ & $ > 29 $ &1014 & $ > $ 45 
& 108 \\
$\phantom{1^+_1 \rightarrow} ~3^+_2$ & &  & $ > $ 0.15  & 0.04  \\
$\phantom{1^+_1 \rightarrow} ~2^+_1$ &$ > $ 0.66 & 4.9 & $ > $ 0.17 & 3.8  \\
$\phantom{1^+_1 \rightarrow} ~3^+_1$ & &  &$ > $ 0.51  & 16  \\
$1^+_2 \rightarrow 1^+_1$ & & 0.003 & & 0.85 \\
$\phantom{1^+_2 \rightarrow} ~2^+_2$ & & 19 & & 1.7  \\
$\phantom{1^+_2 \rightarrow} ~3^+_2$ & &  &$ > $ 0.09  & 11  \\
$\phantom{1^+_2 \rightarrow} ~2^+_1$ &$ > $ 0.09 & 425 & & 7.1  \\
$\phantom{1^+_2 \rightarrow} ~3^+_1$ & &  & & 2  \\
\end{tabular}
\end{table}

\begin{references}
\bibitem{KB} T. T. S. Kuo and G. E. Brown, Nucl. Phys. {\bf 85}, 40 (1966).
\bibitem{HJ} T. Hamada and I. D. Johnston, Nucl. Phys. {\bf 34}, 382 (1962).
\bibitem{Kuo96} T. T. S. Kuo, in {\it New Perspectives in Nuclear Structure},
Proceedings of the Fifth International Spring Seminar on Nuclear Physics,
Ravello, 1995, edited by A. Covello (World Scientific, Singapore, 1996),
p. 159.
\bibitem{Machl94} R. Machleidt and G. Q. Li, Phys. Rep. {\bf 242}, 5 (1994).
\bibitem{Eng95} T. Engeland, M. Hijorth-Jensen, A. Holt, and E. Osnes, Phys.
Scr. {\bf T56}, 58 (1995).
\bibitem{Andr96} F. Andreozzi {\it et al.}, Phys. Rev. C {\bf 54}, 1636 (1996). 
\bibitem{Andr97} F. Andreozzi {\it et. al}, Phys. Rev. C {\bf 56}, R16 (1997).
\bibitem{Holt97} A. Holt {\it et al.}, Nucl. Phys. A {\bf 618}, 107 (1997).
\bibitem{Machl87} R. Machleidt, K. Holinde, and Ch. Elster, Phys. Rep. {\bf149},
1 (1987).
\bibitem{Lipo96} M. Lipoglav\u{s}ek {\it et al.}, Z. Phys A {\bf 356},
239 (1996).
\bibitem{Cov97} A. Covello {\it et al.}, Prog. Part. Nucl. Phys. {\bf 38}, 165 (1997). 
\bibitem{Omtvedt95} P. Omtvedt {\it et al.}, Phys. Rev. Lett. {\bf 75},
3090 (1995).
\bibitem{Zhang96} C. T. Zhang {\it et al.}, Phys. Rev. Lett. {\bf 77},
3743 (1996).
\bibitem{Kerek72} A. Kerek, G. B. Holm, P. Carl\'{e}, and J. McDonald , 
Nucl. Phys.  {\bf A195}, 159 (1972).
\bibitem{Stone89} C. A. Stone, S. H. Faller, and W. B. Walters, Phys. Rev. C
 {\bf 39}, 1963 (1989).
\bibitem{Mach95} H. Mach {\it et al.}, Phys. Rev. C {\bf 51}, 500 (1995).
\bibitem{Sanchez} M. Sanchez-Vega {\it et al.}, Phys. Rev. Lett. {\bf 80},
5504 (1998).
\bibitem{sigma} We define $\sigma= \{ (1/N_d) \sum_i [E_{\rm expt}(i)-
E_{\rm calc}(i)]^2 \}^{1\over2}$, where $N_d$ is the number of data.
\bibitem{Erok94} K. I. Erokhina and V. I. Isakov, Phys. At. Nucl.
{\bf 57}, 198 (1994).
\bibitem{Jiang92} M. F.Jiang, R. Machleidt, D. B. Stout, and T. T. S. Kuo,
Phys Rev. C {\bf 46}, 910 (1992).


\end{references}
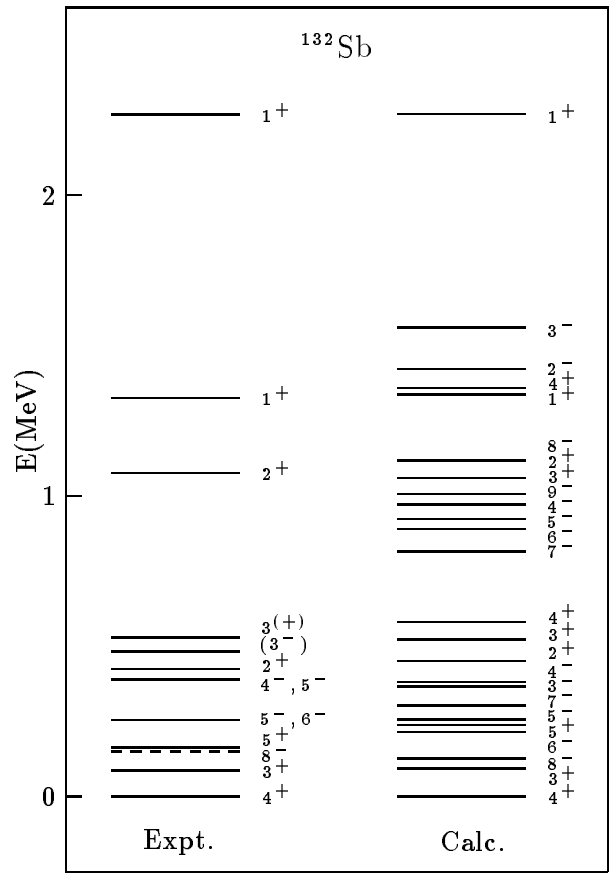
\end{document}